\documentclass[prd,aps,showpacs,nofootinbib,superscriptaddress,notitlepage]{revtex4-1}
\usepackage{epsfig}
\usepackage{amssymb}
\usepackage{amsmath}
\usepackage{graphicx}
\usepackage{psfrag}

\begin{document}

\title{On the linearly polarized gluon distributions in the color dipole model}
\author{Fabio Dominguez}
\affiliation{Department of Physics, Columbia University, New York,
NY, 10027, USA}
\author{Jian-Wei Qiu}
\affiliation{Physics Department, Brookhaven National Laboratory, Upton, NY
11973} \affiliation{C.N.~Yang Institute for Theoretical Physics, Stony Brook
University, Stony Brook, NY 11794, USA}

\author{Bo-Wen Xiao}
\affiliation{Department of Physics, Pennsylvania State University,
University Park, PA 16802, USA}

\author{Feng Yuan}
\affiliation{Nuclear Science Division, Lawrence Berkeley National Laboratory, Berkeley,
CA 94720, USA}

\begin{abstract}
We show that the linearly polarized gluon distributions appear in
the color dipole model as we derive the full cross sections of the
DIS dijet production and the Drell-Yan dijet ($\gamma^*$ jet
correlation) process. Together with the normal
Weizs\"{a}cker-Williams gluon distribution, the linearly polarized
one will contribute to the DIS dijet production cross section
as the coefficient of the $\cos\left( 2 \Delta \phi\right) $ term in the
correlation limit. We also derive the exact results for the cross
section of the Drell-Yan dijet process, and find that the linearly
polarized dipole gluon distribution which is identical to the
normal dipole gluon distribution involves in the cross
section. The results obtained in this paper agree with the
previous transverse momentum dependent factorization study. We further derive the
small-$x$ evolution of these linearly polarized gluon
distributions and find that they rise as $x$ gets small at high
energy.
\end{abstract}

\pacs{24.85.+p, 12.38.Bx, 12.39.St, 13.88.+e}
\maketitle

\section{Introduction}

In small-$x$ physics, two different unintegrated gluon
distributions \cite{Catani:1990eg, Collins:1991ty} (also known as
transverse momentum dependent gluon distributions), namely the
Weizs\"{a}cker-Williams gluon distribution
$xG^{(1)}$ \cite{Kovchegov:1998bi,McLerran:1998nk} and the dipole
gluon distribution $xG^{(2)}$, have been widely used in
the literature. The Weizs\"{a}cker-Williams gluon distribution
can be interpreted as the number density of gluons inside dense
hadrons in light-cone gauge. The dipole gluon distribution,
despite of lacking the probabilistic interpretation, has been
thoroughly studied since it appears in many physical
processes \cite{Iancu:2003xm, Kharzeev:2003wz} and it is defined
via the Fourier transform of the simple color dipole amplitude.
This dipole gluon distribution can be probed directly in
photon-jet correlations and Drell-Yan dijet measurement in pA
collisions. Recent studies \cite{Dominguez:2010xd,
Dominguez:2011wm} on the Weizs\"{a}cker-Williams gluon
distribution indicate that it can be directly measured in DIS
dijet production and its operator definition is related to color
quadrupoles instead of normal color dipoles. Other more
complicated dijet processes in pA collisions (e.g., $qg$ or $gg$
dijets) involve both of these gluon distributions through
convolution in transverse momentum space. The complete calculations were performed in
Ref.~\cite{Dominguez:2010xd, Dominguez:2011wm} in both the Transverse Momentum
Dependent (TMD) factorization formalism and the color dipole
model. The results demonstrate the complete agreement between
these two formalisms in the kinematical region where they are both valid.

Linearly polarized gluon distributions, denoted as
$h_{\perp}^{(i)}(x,q_\perp)$\footnote{Normally it is denoted as
$h^{\perp g}_1(x,q_\perp)$. Here throughout the paper, in order to
avoid confusion on the notation, we use
$h_{\perp}^{(i)}(x,q_\perp)$ with $i=1,2$ to represent the
linearly polarized gluon distributions.},
where $x$ and $q_\perp$ are the active gluon's longitudinal momentum fraction and its transverse momentum, respectively,
were first introduced in Ref.~\cite{Mulders:2000sh}.  This new gluon distribution effectively
measures an averaged quantum interference between a scattering amplitude with an active gluon  polarized along the $x$(or $y$)-axis and a complex conjugate amplitude with an active gluon polarized along the $y$(or $x$)-axis inside an unpolarized hadron.  Because of the unique transverse spin correlation between the two gluon fields of the distribution, the linearly polarized gluon distribution can contribute to a physical observable with $\cos(2\Delta\phi)$-type azimuthal angular dependence, or the azimuthally symmetric observables if they come in pairs.
As proposed in Ref.~\cite{Boer:2010zf}, the linearly polarized gluon
distributions can be directly probed in dijet and heavy quark pair
production processes in electron-hadron collisions.
As expected,
this distribution also
contributes to the cross section in photon pair
productions \cite{Nadolsky:2007ba,Qiu:2011ai} and the Standard Model Higgs boson
production \cite{Catani:2010pd,Sun:2011iw, Boer:2011kf} in hadron-hadron collisions.
Since the integrated parton distributions for incoming
protons were used in the calculations of pA collisions in
Ref.~\cite{Dominguez:2010xd, Dominguez:2011wm}, the linearly
polarized gluon distribution does not enter the cross section
except for the Drell-Yan dijet processes as we show in the later
discussion.

In Ref.~\cite{Metz:2011wb}, the linearly polarized partner of both
Weizs\"{a}cker-Williams and dipole gluon distributions inside an
unpolarized nucleus target is studied in Color Glass Condensate
(CGC) formalism. The corresponding cross sections of deep
inelastic scattering (DIS) dijet production and the Drell-Yan
processes in pA collisions are computed in terms of the TMD
formalism. In both processes, the linearly polarized gluon
distributions appear as the coefficients of the $\cos (2 \Delta
\phi )$ term in the cross section, where they were found to be
consistent with the small-$x$ formalism as
well~\cite{Metz:2011wb}.

Inspired by Ref.~\cite{Metz:2011wb}, we perform the detailed
calculation in the color dipole model for the DIS dijet production
and the Drell-Yan dijet processes in pA collisions and we find
identical results as those with the TMD formalism for the cross
sections in the correlation limit, which is defined as a limit
when the final state dijets are almost back-to-back. For the DIS
dijet production, the complete analysis of the quadrupole
amplitude shows that the linearly polarized gluon distribution of
the Weizs\"{a}cker-Williams type comes from the off-diagonal
expansion of the quadrupole amplitude. Using a hybrid
factorization, we obtain the exact results for the cross section
of the Drell-Yan processes in pA collisions. In the correlation
limit, this exact result reduces to the TMD cross section obtained
in Ref.~\cite{Metz:2011wb}.

Another objective of this paper is to study the small-$x$ evolution
of the linearly polarized gluon distributions. The small-$x$
evolution of the dipole type linearly polarized gluon distribution
is essentially the evolution of the dipole amplitude, which is governed by
the Balitsky-Kovchegov
equation \cite{Balitsky:1995ub+X,Kovchegov:1999yj}. Derived from
the evolution of quadrupoles, the evolution of the linearly
polarized Weizs\"{a}cker-Williams gluon distribution is quite
complicated. Nevertheless, in the dilute regime, we find that both linearly polarized gluon distributions receive the
exponential enhancement in terms of rapidity at high energy as the
normal unpolarized gluon distributions do due to the small-$x$
evolution.

The rest of the paper is organized as follows. In Sec. II, we
calculate the cross sections of the DIS dijet production and the
Drell-Yan processes in pA collisions and demonstrate that the
linearly polarized gluon distributions naturally arise in the
dipole model.  We discuss the small-$x$ evolution equations of the
linearly polarized gluon distributions in Sec. III. The summary
and further discussions are given in Sec. IV.

\section{The linearly polarized gluon distribution in Dipole model}

In this section, following Ref.~\cite{Dominguez:2011wm}, we show that
the cross section of the DIS dijet production and the Drell-Yan dijet
process in the color dipole model, namely the CGC approach, involves the
linearly polarized gluon distribution as well. The reason why this does not appear in the original work in \cite{Dominguez:2011wm} is that there the azimuthal orientation of the outgoing partons was averaged over.

\subsection{DIS dijet production}

After averaging over the photon's polarization and summing over the quark
and antiquark helicities and colors, the cross section of the DIS dijet
production in the color dipole model can be cast into
\begin{eqnarray}
\frac{d\sigma ^{\gamma_{T,L}^{\ast }A\rightarrow q\bar{q}X}}{d^3k_1d^3k_2}
&=&N_{c}\alpha _{em}e_{q}^{2}\delta(p^+-k_1^+-k_2^+) \int \frac{\text{d}%
^{2}x_1}{(2\pi)^{2}}\frac{\text{d}^{2}x_1^{\prime }}{(2\pi )^{2}} \frac{%
\text{d}^{2}x_2}{(2\pi)^{2}}\frac{\text{d}^{2}x_2^{\prime }}{(2\pi )^{2}}
\notag \\
&&\times e^{-ik_{1\perp }\cdot(x_1-x_1^{\prime })} e^{-ik_{2\perp }\cdot
(x_2-x_2^{\prime })} \sum_{\lambda\alpha\beta} \psi_{\alpha\beta}^{T, L
\lambda}(x_1-x_2) \psi_{\alpha\beta}^{T, L\lambda*}(x_1^{\prime
}-x_2^{\prime })  \notag \\
&&\times \left[1+S^{(4)}_{x_g}(x_1,x_2;x_2^{\prime },x_1^{\prime})
-S^{(2)}_{x_g}(x_1,x_2)-S^{(2)}_{x_g}(x_2^{\prime },x_1^{\prime })\right] \ ,
\label{xsdis}
\end{eqnarray}
where the two- and four-point functions, which are characterized by the
Wilson lines, take care of the multiple scatterings between the $q\bar q$-pair
and the target. They are defined as
\begin{eqnarray}
&&S_{x_g}^{(2)}(x_1,x_2)=\frac{1}{N_c}\left\langle\text{Tr}%
U(x_1)U^\dagger(x_2)\right\rangle_{x_g}\ , \\
&&S_{x_g}^{(4)}(x_1,x_2;x_2^{\prime},x_1^{\prime})=\frac{1}{N_c}\left\langle
\text{Tr}U(x_1)U^\dagger(x_1^{\prime})
U(x_2^{\prime})U^\dagger(x_2)\right\rangle_{x_g}\ , \\
\text{with} \quad && U(x)=\mathcal{P}\exp\left\{ig_S\int_{-\infty}^{+\infty}
\text{d}x^+\,T^cA_c^-(x^+,x)\right\}.
\end{eqnarray}
The notation $\langle\dots\rangle_{x_g}$ is used for the CGC average of the
color charges over the nuclear wave function where $x_g$ is the smallest
fraction of longitudinal momentum probed, and is determined by the
kinematics. The splitting wave function of the virtual photon with
longitudinal momentum $p^+$ and virtuality $Q^2$ in transverse coordinate
space takes the form,
\begin{align}
\psi^{T\,\lambda}_{\alpha\beta}(p^+,z,r)&=2\pi\sqrt{\frac{2}{p^+}}
\begin{cases}
i\epsilon_fK_1(\epsilon_f|r|)\tfrac{r\cdot\epsilon^{(1)}_\perp}{|r|}
[\delta_{\alpha+}\delta_{\beta+}(1-z)+\delta_{\alpha-}\delta_{\beta-}z], &
\lambda=1, \\
i\epsilon_fK_1(\epsilon_f|r|)\tfrac{r\cdot\epsilon^{(2)}_\perp}{|r|}
[\delta_{\alpha-}\delta_{\beta-}(1-z)+\delta_{\alpha+}\delta_{\beta+}z], &
\lambda=2,%
\end{cases}
\\
\psi^L_{\alpha\beta}(p^+,z,r)&=2\pi\sqrt{\frac{4}{p^+}}z(1-z)QK_0(
\epsilon_f|r|)\delta_{\alpha\beta}.
\end{align}
where $z$ is the momentum fraction of the photon carried by the quark, $\lambda$ is the photon polarization, $\alpha$ and $\beta$ are the quark and
antiquark helicities, $r$ the transverse separation of the pair, $
\epsilon_f^2=z(1-z)Q^2$, and the quarks are assumed to be massless.

In order to take the correlation limit, we introduce the transverse
coordinate variables: $u=x_1-x_2$ and $v=zx_1+(1-z)x_2$, and similarly for
the primed coordinates, with respective conjugate momenta $\tilde
P_\perp=(1-z)k_{1\perp}-zk_{2\perp}$\footnote{One could also define $v=\frac{1}{2}(x_1+x_2)$ in this process since the
virtual photon does not have initial interactions with the nucleus target,
then the respective conjugate momentum is $P_\perp=\frac{1}{2}
(k_{1\perp}-k_{2\perp})\simeq \tilde P_\perp$. $P_\perp$ is the relative
momentum of outgoing partons respect to the center of mass frame.
Nevertheless, the following derivation remains the same in this case.} and $q_\perp$. The correlation limit ($\tilde P_\perp\simeq
k_{1\perp}\simeq k_{2\perp} \gg q_\perp$) is therefore enforced by assuming $u$ and $u^{\prime}$ are small as compared to $v$ and $v^\prime$ and then expanding
the integrand with respect to these two variables before performing the
Fourier transform. Following the derivation in Ref.~\cite{Dominguez:2011wm},
one can find that the lowest order expansion in $u$ and $u^{\prime }$ of the
last line of Eq.~(\ref{xsdis}) gives
\begin{equation}
-u_{i}u_{j}^{\prime }\frac{1}{N_{c}}\langle \text{Tr}\left[ \partial^{i}U(v)\right] U^{\dagger }(v^{\prime })\left[ \partial ^{j}U(v^{\prime })\right]
U^{\dagger }(v)\rangle _{x_{g}}\ .
\end{equation}
With the help of the identities
\begin{eqnarray}
\int \frac{d^2 u}{(2\pi)^2} \frac{d^2 u^{\prime}}{(2\pi)^2}e^{-i\tilde{P}
_{\perp}\cdot (u-u^{\prime})}u_{i}u_{j}^{\prime }
\nabla_uK_0(\epsilon_fu)\cdot \nabla_{u^{\prime}}K_0(\epsilon_fu^{\prime})
&=&\frac{1}{(2\pi)^2}\left[\frac{\delta_{ij}}{(\tilde{P}_{\perp}^2+
\epsilon_f^2)^2}-\frac{4\epsilon_f^2\tilde{P}_{\perp i}\tilde{P}_{\perp j}}{(
\tilde{P}_{\perp}^2+\epsilon_f^2)^4}\right]\, , \\
\int \frac{d^2 u}{(2\pi)^2} \frac{d^2 u^{\prime}}{(2\pi)^2}e^{-i\tilde{P}
_{\perp}\cdot (u-u^{\prime})}u_{i}u_{j}^{\prime }
K_0(\epsilon_fu)K_0(\epsilon_fu^{\prime}) &=&\frac{1}{(2\pi)^2}\frac{4\tilde{
P}_{\perp i}\tilde{P}_{\perp j}}{(\tilde{P}_{\perp}^2+\epsilon_f^2)^4}\, ,
\end{eqnarray}
one can integrate over $u$ and $u^{\prime }$ and obtain the
complete differential cross section in the correlation limit,
\begin{eqnarray}
\frac{d\sigma ^{\gamma _{T}^{\ast }A\rightarrow q\bar{q}X}}{d\mathcal{P.S.}}
&=&\alpha _{em}e_{q}^{2}\alpha _{s}\delta \left( x_{\gamma ^{\ast
}}-1\right) z(1-z)\left( z^{2}+(1-z)^{2}\right) \left[\frac{\delta_{ij}}{(
\tilde{P}_{\perp}^2+\epsilon_f^2)^2}-\frac{4\epsilon_f^2\tilde{P}_{\perp i}
\tilde{P}_{\perp j}}{(\tilde{P}_{\perp}^2+\epsilon_f^2)^4}\right]  \notag \\
&&\times (16\pi ^{3})\int \frac{d^{3}vd^{3}v^{\prime }}{(2\pi )^{6}}
e^{-iq_{\perp }\cdot (v-v^{\prime })}2\left\langle \text{Tr}\left[ F^{i-}({v}
)\mathcal{U}^{[+]\dagger }F^{j-}({v}^{\prime })\mathcal{U}^{[+]}\right]
\right\rangle _{x_{g}}\ ,  \label{dipoledis} \\
\frac{d\sigma ^{\gamma _{L}^{\ast }A\rightarrow q\bar{q}X}}{d\mathcal{P.S.}}
&=&\alpha _{em}e_{q}^{2}\alpha _{s}\delta \left( x_{\gamma ^{\ast
}}-1\right)4 z^{2}(1-z)^{2}\frac{4\epsilon_f^2\tilde{P}_{\perp i}\tilde{P}
_{\perp j}}{(\tilde{P}_{\perp}^2+\epsilon_f^2)^4}  \notag \\
&&\times (16\pi ^{3})\int \frac{d^{3}vd^{3}v^{\prime }}{(2\pi )^{6}}
e^{-iq_{\perp }\cdot (v-v^{\prime })}2\left\langle \text{Tr}\left[ F^{i-}({v}
)\mathcal{U}^{[+]\dagger }F^{j-}({v}^{\prime })\mathcal{U}^{[+]}\right]
\right\rangle _{x_{g}}\ .  \label{dipoledisl}
\end{eqnarray}
Here we have used the identity
\begin{equation}
-\langle \text{Tr}\left[ \partial _{i}U(v)\right] U^{\dagger }(v^{\prime })
\left[ \partial _{j}U(v^{\prime })\right] U^{\dagger }(v)\rangle
_{x_{g}}=g_{S}^{2}\int_{-\infty }^{\infty }\text{d}v^{+}\text{d}v^{\prime
+}\left\langle \text{Tr}\left[ F^{i-}({v})\mathcal{U}^{[+]\dagger
}F^{j-}({v}^{\prime })\mathcal{U}^{[+]}\right] \right\rangle _{x_{g}}\ ,
\end{equation}
where the gauge link $\mathcal{U}^{[+]}$ connects the two coordinate points by means of longitudinal gauge links going to $+\infty$ and a transverse link at infinity which does not contribute when the appropriate boundary conditions are taken.

If one integrates over the orientation of $\tilde P_{\perp}$, one can
replace $\tilde P_{\perp i}\tilde P_{\perp j}$ by $\frac{1}{2}\delta_{ij}
\tilde P^2_{\perp}$.\footnote{In the derivation of Ref.~\cite{Dominguez:2011wm}, we have employed this as
an underlying assumption.} This replacement allows us to reduce the above
expressions into Eqs.~(30) and (31) in Ref.~\cite{Dominguez:2011wm} which
only involve the conventional Weizs\"{a}cker-Williams gluon distribution.

Now we are ready to show that the linearly polarized Weizs\"{a}cker-Williams
gluon distribution can also arise naturally in the color dipole model. Since
the indices $i,\,j$ are symmetric, we can decompose the operator expression
appearing in Eqs~(\ref{dipoledis}) and (\ref{dipoledisl}) into two parts with
one part involving only $\delta _{ij}$ and the other part being traceless,
\begin{eqnarray}
&&4\int \frac{d^{3}vd^{3}v^{\prime }}{(2\pi )^{3}}e^{-iq_{\perp }\cdot
(v-v^{\prime })}\left\langle \text{Tr}\left[ F^{i-}({v})\mathcal{U}%
^{[+]\dagger }F^{j-}({v}^{\prime })\mathcal{U}^{[+]}\right] \right\rangle
_{x_{g}}  \notag \\
&=&\frac{1}{2}\delta ^{ij}xG^{(1)}(x,q_{\perp })+\frac{1}{2}\left( \frac{
2q_{\perp }^{i}q_{\perp }^{j}}{q_{\perp }^{2}}-\delta ^{ij}\right) xh_{\perp
}^{(1)}(x,q_{\perp }).
\label{eq:decomposition}
\end{eqnarray}%
Here $xG^{(1)}(x,q_{\perp })$ is the conventional Weizs\"{a}cker-Williams
gluon distribution while the coefficient of the traceless tensor $xh_{\perp
}^{(1)}(x,q_{\perp })$ is the so-called linearly polarized partner of the
conventional Weizs\"{a}cker-Williams gluon distribution.

The physical meaning or interpretation of these two gluon distributions can
be better represented in a frame in which the two components of the
transverse momentum $q_\perp^j$ with $j=1,2$ or $j=x,y$ are the same.
With $q_\perp^{x}=q_\perp^{y}$ in this frame, the two symmetric projection
operators in Eq.~(\ref{eq:decomposition}) can be written as,
\begin{eqnarray}
\frac{1}{2}\delta^{ij}
&=&
\frac{1}{2}
 \begin{pmatrix}
      1 & 0 \\
      0 & 1
    \end{pmatrix}
=\frac{1}{2}
\left(
e_x^i e_x^j + e_y^i e_y^j \right)
=\frac{1}{2}
\left[ \varepsilon_+^{*i} \varepsilon_+^{j} + \varepsilon_-^{*i} \varepsilon_-^{j} \right]
\, ,
\label{eq:diagonal}\\
\frac{1}{2}\left(
\frac{2q_{\perp }^{i}q_{\perp }^{j}}{q_{\perp }^{2}}
-\delta^{ij}\right)
&=&
\frac{1}{2}
 \begin{pmatrix}
      0 & 1 \\
      1 & 0
    \end{pmatrix}
=\frac{1}{2}
\left(
e_x^i e_y^j + e_y^i e_x^j \right)
=\frac{1}{2i}
\left[ \varepsilon_+^{*i} \varepsilon_-^{j} - \varepsilon_-^{*i} \varepsilon_+^{j} \right]
\, ,
\label{eq:offdiagonal}
\end{eqnarray}
where $e_x^i=(1,0)$ and $e_y^i=(0,1)$ are 2-dimensional unit vectors
along $x$-axis and $y$-axis, respectively, which could be interpreted as
two orthogonal {\em linear} polarization vectors for transversely polarized gluons.
As shown in Eqs.~(\ref{eq:diagonal}) and (\ref{eq:offdiagonal}),
these two symmetric projection operators can also be expressed
in terms of the two orthogonal {\em circular} polarization vectors for transversely
polarized gluons, $\varepsilon_{\pm}^j \equiv  [\mp e_x^j - i\, e_y^j]/\sqrt{2}$.
For the comparison, we also list here the antisymmetric projection operator
for the polarized gluon helicity distribution,
\begin{equation}
\frac{1}{2}\left( i \epsilon_\perp^{ij}\right)
=
\frac{1}{2}
 \begin{pmatrix}
      0 & i \\
      -i & 0
    \end{pmatrix}
=\frac{1}{2}\, i
\left(
e_x^i e_y^j - e_y^i e_x^j \right)
=\frac{1}{2}
\left[ \varepsilon_+^{*i} \varepsilon_+^{j} - \varepsilon_-^{*i} \varepsilon_-^{j} \right]
\, .
\label{eq:pol}
\end{equation}
From Eqs.~(\ref{eq:diagonal}) and (\ref{eq:pol}), it is natural to interpret $G^{(1)}$
as a probability distribution to find unpolarized gluons, while the polarized gluon helicity
distribution could be interpreted as a {\em difference} of two probability distributions to
find positive helicity gluons and negative helicity gluons, respectively. From Eq.~(\ref{eq:offdiagonal}), it appears that $h_\perp^{(1)}$ does not have a
probability interpretation in terms of the base polarization vectors $\varepsilon_{\pm}^j$, which are the eigenstates of angular momentum operators. \footnote{However, if one chooses different base polarization vectors as $e_1^i=\frac{1}{\sqrt{2}}(1,1)$ and $e_2^i=\frac{1}{\sqrt{2}}(1,-1)$, which are not the eigenstates of angular momentum operators, one can find that Eq.~(\ref{eq:offdiagonal}) becomes $\frac{1}{2}\left(
e_1^i e_1^j - e_2^i e_2^j \right)$ which would allow us to interpret $h_\perp^{(1)}$ as the linearly polarized gluon density along the direction of the linear polarization. In a general frame, the polarization vectors are found to be $e_1^i=(\cos \phi,\sin \phi)$ and $e_2^i=(\sin \phi,-\cos \phi)$ which convert Eq.(\ref{eq:proj_h}) into $\frac{1}{2}\left(e_1^i e_1^j - e_2^i e_2^j \right)$ as well. This indicates that the interpretation of the linearly polarized gluon distributions depends on the choice of the polarization vectors.} Instead, it could be interpreted as a transverse spin
correlation function to find the gluons in the amplitude and complex conjugate
amplitude  to be in two orthogonal polarization states.  In a general frame,
$q_\perp^j=(q_\perp^x,q_\perp^y)=q_\perp(\cos\phi,\sin\phi)$,
the projection operator for $h_\perp^{(1)}$ can be written as,
\begin{equation}
\frac{1}{2}\left(
\frac{2q_{\perp }^{i}q_{\perp }^{j}}{q_{\perp }^{2}}
-\delta^{ij}\right)
=
\frac{1}{2}
    \begin{pmatrix}
      \cos(2\phi)        &     \sin(2\phi) \\
      \sin(2\phi)        &   - \cos(2\phi)
    \end{pmatrix}\, ,
\label{eq:proj_h}
\end{equation}
which includes the special case in Eq.~(\ref{eq:offdiagonal})
when $\phi=\pi/4$.  Since the projection operator in Eq.~(\ref{eq:proj_h})
is proportional to a rotation matrix of the azimuthal angle, the $h_\perp^{(1)}$
could also be interpreted as ``azimuthal correlated'' gluon distributions
\cite{Nadolsky:2007ba,Catani:2010pd}. Because the gluons in the amplitude and
complex conjugate amplitude are in different transverse spin states,
this kind of gluon distributions could contribute to the observables
with $\cos(2\Delta\phi)$-type azimuthal dependence, or azimuthal symmetric
observables if they come in pairs.

Substitute Eq.~(\ref{eq:decomposition}) into
Eqs~(\ref{dipoledis}) and (\ref{dipoledisl}), we obtain
\begin{eqnarray}
\frac{d\sigma ^{\gamma _{T}^{\ast }A\rightarrow q\bar{q}X}}{d\mathcal{P.S.}}
&=&\alpha _{em}e_{q}^{2}\alpha _{s}\delta \left( x_{\gamma ^{\ast
}}-1\right) z(1-z)\left( z^{2}+(1-z)^{2}\right) \frac{\epsilon _{f}^{4}+%
\tilde{P}_{\perp }^{4}}{(\tilde{P}_{\perp }^{2}+\epsilon _{f}^{2})^{4}}
\notag \\
&&\times \left[ xG^{(1)}(x,q_{\perp })-\frac{2\epsilon _{f}^{2}\tilde{P}%
_{\perp }^{2}}{\epsilon _{f}^{4}+\tilde{P}_{\perp }^{4}}\cos  \left(2\Delta \phi\right)xh_{\perp }^{(1)}(x,q_{\perp })\right] ,
\label{dipoledis2} \\
\frac{d\sigma ^{\gamma _{L}^{\ast }A\rightarrow q\bar{q}X}}{d\mathcal{P.S.}}
&=&\alpha _{em}e_{q}^{2}\alpha _{s}\delta \left( x_{\gamma ^{\ast
}}-1\right) z^{2}(1-z)^{2}\frac{8\epsilon _{f}^{2}\tilde{P}_{\perp }^{2}}{(%
\tilde{P}_{\perp }^{2}+\epsilon _{f}^{2})^{4}}  \notag \\
&&\times \left[ xG^{(1)}(x,q_{\perp })+\cos \left(2\Delta \phi\right)xh_{\perp }^{(1)}(x,q_{\perp })\right] \ .
\label{dipoledisl2}
\end{eqnarray}
where $\Delta \phi=\phi _{\tilde{P}_{\perp }}-\phi _{q_{\perp }}$ with $\phi _{\tilde{P}_{\perp }}$ and $\phi _{q_{\perp }}$ being the
azimuthal angle of $\tilde{P}_{\perp }$ and $q_{\perp }$, respectively. This
result is in complete agreement with the one obtained in Ref.~\cite{Metz:2011wb} by using the TMD approach. The coefficient of the $\cos  \left(2\Delta \phi\right)$ term in the above cross section
can provide us the direct information of the linearly polarized Weizs\"{a}cker-Williams gluon distribution $xh_{\perp }^{(1)}(x,q_{\perp })$. It is
also easy to see that the $xh_{\perp }^{(1)}(x,q_{\perp })$ term vanishes if
one averages the cross section over the orientation of either $\tilde{P}_{\perp }$ or $q_{\perp }$ due to the factor $\cos  \left(2\Delta \phi\right)$. This is transparent when one uses the variables $P_{\perp }$ and $q_{\perp }$ since they can be interpreted as the relative
transverse momentum with respect to the center of mass frame of these two outgoing
partons and the total transverse momentum of the CM frame, respectively.

Last but not least, one can see that the contribution from the linearly polarized
gluon distribution vanishes if $Q=0$, i.e., the real photon nucleus
scattering only involves the conventional Weizs\"{a}cker-Williams gluon
distribution.
This is because the real photon cannot generate a $\cos\left(2\Delta\phi\right)$-type
transverse spin correlation that matches the transverse spin correlation
generated by $h_\perp^{(1)}$.

Let us now study the behavior of $xh_{\perp }^{(1)}(x,q_{\perp })$ in the McLerran-Venugopalan (MV) model\cite{McLerran:1993ni}
for a large nucleus with $A$ nucleons inside. Using the quadrupole results calculated in Ref.~\cite{Dominguez:2011wm}, one can cast the analytical form of $xh_{\perp }^{(1)}(x,q_{\perp })$ into\cite{Metz:2011wb}
\begin{eqnarray}
xh_{\perp }^{(1)}(x,q_{\perp }) &=&\frac{2}{\alpha _{s}}\left( \delta ^{ij}-%
\frac{2q_{\perp }^{i}q_{\perp }^{j}}{q_{\perp }^{2}}\right) \int \frac{%
d^{2}vd^{2}v^{\prime }}{(2\pi )^{2}(2\pi )^{2}}e^{-iq_{\perp }\cdot
(v-v^{\prime })}\langle \text{Tr}\left[ \partial _{i}U(v)\right] U^{\dagger
}(v^{\prime })\left[ \partial _{j}U(v^{\prime })\right] U^{\dagger
}(v)\rangle _{x_{g}}  \notag \\
&=&\frac{S_{\perp }}{2\pi ^{3}\alpha _{s}}\frac{N_{c}^{2}-1}{N_{c}}
\int_{0}^{\infty }dr_{\perp }r_{\perp }\frac{J_{2}(q_{\perp }r_{\perp })}{
r_{\perp }^{2}\ln \frac{1}{r_{\perp }^{2}\Lambda ^{2}}}\left[ 1-\exp \left( -
\frac{1}{4}r_{\perp }^{2}Q_{sg}^{2}\right) \right] ,
\end{eqnarray}%
where $J_{2}(q_{\perp }r_{\perp })$ is the Bessel function of the first kind
and $Q_{sg}^{2}=\alpha _{s}g^{2}N_{c}\mu ^{2}\ln \frac{1}{r_{\perp
}^{2}\Lambda ^{2}}$ with $\mu ^{2}=\frac{A}{2S_{\perp }}$. For $q_{\perp
}^{2}\gg Q_{sg}^{2}$, we find that $xh_{\perp }^{(1)}(x,q_{\perp })\simeq
\frac{\alpha _{s}AC_{F}N_{c}}{\pi ^{2}q_{\perp }^{2}}$ which is identical to
$xG^{(1)}(x,q_{\perp })$ and agrees with the perturbative QCD results. It is
important to notice that it scales like $A$ since each nucleon contributes additively in the dilute regime. In this regime, the dominant contribution to the gluon distribution comes from a single two-gluon exchange with a transverse momentum transfer $q_\perp$ in the color dipole picture. For the case $\Lambda ^{2}\ll q_{\perp
}^{2}\ll Q_{sg}^{2}$ one absorbs the $\ln \frac{1
}{r_{\perp }^{2}\Lambda ^{2}}$ factor into the definition of the saturation
momentum and finds $xh_{\perp }^{(1)}(x,q_{\perp })\simeq \frac{\alpha
_{s}AC_{F}N_{c}}{\pi ^{2}Q_{sg}^{2}}$ which is an approximate constant. It
scales like $A^{2/3}$ since $Q_{sg}^{2}\sim A^{1/3}$ as a result of strong
nuclear shadowing. It is interesting to note that, in the low $q_{\perp }^{2}$ region, the effect of multiple scatterings between probes and target nuclei can be viewed as or attributed to a single scattering with the momentum transfer of order $Q_{sg}^{2}$. As compared to the small $q_{\perp }^{2}$ behavior of the
conventional Weizs\"{a}cker-Williams gluon distribution $xG^{(1)}(x,q_{\perp
})\simeq \frac{S_{\perp }}{4\pi ^{3}\alpha _{s}}\frac{N_{c}^{2}-1}{N_{c}}\ln
\frac{Q_{sg}^{2}}{q_{\perp }^{2}}$, we find that $\frac{xG^{(1)}(x,q_{\perp })
}{xh_{\perp }^{(1)}(x,q_{\perp })}\simeq \ln \frac{q_{\perp }^{2}}{\Lambda
^{2}}\ln \frac{Q_{sg}^{2}}{q_{\perp }^{2}}\gg 1$ where we have replaced $r_{\perp }\ $by $\frac{1}{q_{\perp }}$.
These gluon distributions obtained in the MV model can be viewed as an initial condition for the small-$x$ evolution. In addition, we can also find that $xG^{(1)}(x,q_{\perp })\geq xh_{\perp }^{(1)}(x,q_{\perp })$ for any value of $q_\perp$ which ensures the positivity of the total cross section.

\subsection{Drell-Yan dijet process}

Following the prompt photon-jet correlation calculation in Ref.~\cite%
{Dominguez:2011wm}, it is straightforward to calculate the cross section of
dijet ($q\gamma^{\ast}$) production in Drell-Yan processes in pA collisions.
The calculation is essentially the same, except for the slightly
different splitting function since the final state virtual photon, which
eventually decays into a di-lepton pair, has a finite invariant mass $M$. By
taking into account the photon invariant mass, the splitting wave functions
of a quark with longitudinal momentum $p^+$ splitting into a quark and
virtual photon pair in transverse coordinate space become
\begin{eqnarray}
\psi^{T\, \lambda}_{\alpha\beta}(p^+,k_1^+,r)&=&2\pi \sqrt{\frac{2}{k_1^+}}%
\begin{cases}
i\epsilon_MK_1( \epsilon_M|r|)\frac{r\cdot\epsilon^{(1)}_\perp}{|r|}%
(\delta_{\alpha-}\delta_{\beta-}+(1-z)\delta_{\alpha+}\delta_{\beta+}), &
\lambda=1, \\
i\epsilon_MK_1( \epsilon_M|r|)\frac{r\cdot\epsilon^{(2)}_\perp}{|r|}%
(\delta_{\alpha+}\delta_{\beta+}+(1-z)\delta_{\alpha-}\delta_{\beta-}), &
\lambda=2.%
\end{cases}
\ ,  \label{wvfunction} \\
\psi^L_{\alpha\beta}(p^+,k_1^+,r)&=&2\pi\sqrt{\frac{2}{k_1^+}}(1-z)MK_0(
\epsilon_M|r|)\delta_{\alpha\beta},
\end{eqnarray}
where $\epsilon _{M}^{2}=(1-z)M^{2}$, $\lambda$ is the photon polarization, $\alpha,\beta$ are helicities for the
incoming and outgoing quarks, and $z=\frac{k_1^+}{p^+}$ is the momentum fraction of the incoming quark carried by
the photon.

At the end of the day, for the correlation between the final state virtual
photon and quark in $pA$ collisions, we have
\begin{eqnarray}
\frac{d\sigma _{\text{DP}}^{pA\rightarrow \gamma ^{\ast }q+X}}{
dy_{1}dy_{2}d^{2}k_{1\perp }d^{2}k_{2\perp }} &=&
\sum_{f}x_{p}q_{f}(x_{p},\mu )\frac{\alpha _{e.m.}e_{f}^{2}}{2\pi ^{2}}
\left( 1-z\right)z^{2} S_\perp F_{x_{g}}(q_{\perp })  \notag \\
&&\times \left\{ \left[ 1+\left( 1-z\right) ^{2}\right] \frac{q_{\perp }^{2}
}{\left[ \tilde{P}_{\perp }^{2}+\epsilon _{M}^{2}\right] \left[ (\tilde{P}
_{\perp }+zq_{\perp })^{2}+\epsilon _{M}^{2}\right] }\right.  \notag \\
&&\left. -\epsilon _{M}^{2}\left[ \frac{1}{\tilde{P}_{\perp }^{2}+\epsilon
_{M}^{2}}-\frac{1}{(\tilde{P}_{\perp }+zq_{\perp })^{2}+\epsilon _{M}^{2}}
\right] ^{2}\right\} ,  \label{dyc}
\end{eqnarray}%
with $F_{x_g}(q_\perp)=\int \frac{d^2r_\perp}{(2\pi)^2}e^{-iq_\perp\cdot
r_\perp} \frac{1}{N_c}\left\langle\text{Tr}U(0)U^\dagger(r_\perp)\right
\rangle_{x_g}$, $q_\perp=k_{1\perp}+k_{2\perp }$ and $\tilde{P}_{\perp
}=(1-z)k_{1\perp }-zk_{2\perp }$. In the MV
model, $F_{x_g}(q_\perp)\simeq \frac{1}{\pi Q_{sq}^2} \exp \left(-\frac{q_{\perp}^2}{Q_{sq}^2}\right)$ with $Q_{sq}^2=\frac{C_F}{N_c}Q_{sg}^2$ being the quark saturation momentum. $q_{f}(x_{p},\mu )$ is the integrated
quark distribution with flavor $f$ in the proton projectile. Here we used
the hybrid factorization which allows us to use integrated parton
distributions since the proton projectile is considered to be dilute as
compared to the nucleus target. The first term in the curly brackets arises
solely from the transverse splitting function in Eq.~(\ref{wvfunction})
while the second term is the sum of contributions from both the transverse
and longitudinal splitting functions. We would like to emphasize that the
above cross section in Eq.~(\ref{dyc}) is an exact result regardless of the
relative size between $q_\perp$ and $\tilde{P}_{\perp }$. By taking the
correlation limit, namely $q_\perp \ll \tilde{P}_{\perp }$, we arrive at the
result which is identical to the one obtained from TMD factorization \cite{Metz:2011wb} \footnote{To compare with Ref.~\cite{Metz:2011wb}, one can compute the Mandelstam
variables and find that $\hat s=(k_1+k_2)^2=M^2+\frac{(1-z)(M^2+k_{1\perp}^2)}{z}+\frac{zk_{2\perp}^2}{(1-z)} -2k_{1\perp}\cdot k_{2\perp} =\frac{\tilde{P}_{\perp }^{2}+\epsilon
_{M}^{2}}{z(1-z)}$, $\hat u=\frac{\tilde{P}_{\perp }^{2}+\epsilon _{M}^{2}}{z%
}$ and $\hat t=\frac{\tilde{P}_{\perp }^{2}}{1-z}$.}
\begin{eqnarray}
\left .\frac{d\sigma _{\text{DP}}^{pA\rightarrow \gamma ^{\ast }q+X}}{%
dy_{1}dy_{2}d^{2}k_{1\perp }d^{2}k_{2\perp }}\right|_{q_\perp \ll \tilde{P}%
_{\perp }} = \sum_{f}x_{p}q_{f}(x_{p},\mu ) xG^{(2)}(x_g, q_\perp) \left[
H_{qg\to q\gamma^{\ast}} -\cos \left(2\Delta \phi\right) H_{qg\to q\gamma^{\ast}}^{\perp} \right] ,  \label{dyc2}
\end{eqnarray}
with $\Delta \phi=\phi _{\tilde{P}_{\perp }}-\phi _{q_{\perp }}$, $xG^{(2)}(x,q_\perp)= \frac{q_{\perp }^{2}N_{c}}{2\pi^2 \alpha_s}S_{\perp } F_{x_{g}}(q_{\perp })$ and
\begin{eqnarray}
H_{qg\to q\gamma^{\ast}}&=&\frac{\alpha_s \alpha _{e.m.}e_{f}^{2}\left(
1-z\right)z^{2}}{N_c}\left\{\frac{1+\left( 1-z\right) ^{2}}{\left[ \tilde{P}%
_{\perp }^{2}+\epsilon _{M}^{2}\right]^2}-\frac{2z^2\epsilon _{M}^{2} \tilde{%
P}_{\perp }^{2}}{\left[ \tilde{P}_{\perp }^{2}+\epsilon _{M}^{2}\right]^4}%
\right\}, \\
H_{qg\to q\gamma^{\ast}}^{\perp}&=&\frac{\alpha_s \alpha
_{e.m.}e_{f}^{2}\left( 1-z\right)z^{2}}{N_c} \frac{2z^2 \epsilon _{M}^{2}
\tilde{P}_{\perp }^{2}}{\left[ \tilde{P}_{\perp }^{2}+\epsilon _{M}^{2}%
\right]^4}.
\end{eqnarray}
In this case, the relevant gluon distribution is the so-called dipole gluon
distribution as demonstrated in Ref.~\cite{Dominguez:2010xd, Dominguez:2011wm, Bomhof:2006dp}. As discussed
in Ref.~\cite{Metz:2011wb}, according to the operator definition of dipole
type gluon distributions \cite{Dominguez:2010xd, Dominguez:2011wm, Bomhof:2006dp},
\begin{eqnarray}
xG^{ij}_{\text{DP}}(x,q_{\perp }) &=&2\int \frac{d\xi ^{-}d\xi _{\perp }}{%
(2\pi )^{3}P^{+}}e^{ixP^{+}\xi ^{-}-iq_{\perp }\cdot \xi _{\perp }}\langle P|%
\text{Tr}\left[F^{+i}(\xi ^{-},\xi _{\perp })\mathcal{U}^{[-]\dagger
}F^{+j}(0)\mathcal{U}^{[+]}\right]|P\rangle \ , \\
&=& \frac{q_{\perp }^iq_{\perp }^jN_{c}}{2\pi^2 \alpha_s}S_{\perp }
F_{x_{g}}(q_{\perp }) , \\
&=&\frac{1}{2}\delta^{ij} xG^{(2)}(x,q_{\perp})+\frac{1}{2}\left(\frac{%
2q_{\perp}^iq_{\perp }^j}{q_{\perp }^2}-\delta^{ij}\right)xh^{(2)}_{%
\perp}(x,q_{\perp}),\label{TMDdp}
\end{eqnarray}
where the gauge link $\mathcal{U}^{[-]}$ is composed by longitudinal gauge links going to $-\infty$.
This shows that the linearly polarized partner of the dipole gluon
distribution is exactly the same as the dipole gluon distribution\footnote{%
There is a factor of 2 between these two distributions in Ref.~\cite{Metz:2011wb} due to different normalization.}. From Eq.~(\ref{TMDdp}), with the
proper normalization, we can also find that the linearly polarized gluon
distribution $xh^{(2)}_{\perp}(x,q_\perp)=xG^{(2)}(x,q_\perp)$.

Furthermore,
one can see that for the prompt photon-jet correlation, the linearly
polarized gluon distribution does not contribute since $H_{qg\to
q\gamma^{\ast}}^{\perp}$ vanishes when $M=0$.
This is also due to the fact that the real photon in the final state
cannot generate the transverse spin correlation that matches the
transverse spin correlation of the incoming gluon in the $qg\to q\gamma$
subprocess.  It takes two matched transverse spin correlations to get
a nonvanish observable effect.

\subsection{Resummation}
For the purpose of the Collins-Soper-Sterman resummation\cite{Collins:1984kg}
discussed in Ref.~\cite{Sun:2011iw}, it is also useful to define the coordinate
expression of the linearly polarized Weizs\"{a}cker-Williams gluon distribution as follows
\begin{equation}
x\tilde{h}_{\perp}^{(1)ij}(x,b_\perp)=\frac{1}{2}\int \text{d}^{2}q_\perp e^{-iq_\perp \cdot b_\perp} \left( \frac{2q_\perp^i q_\perp^{j}}{%
q_\perp ^{2}}-\delta^{ij}\right) xh_{\perp}^{(1)}(x,q_\perp ),
\end{equation}
and it is straightforward to find that in the MV model
\begin{equation}
x\tilde{h}_{\perp}^{(1)ij}(x,b_\perp) = \frac{1}{2}\left(\delta^{ij}- \frac{2b_\perp^i b_\perp^{j}}{%
b_\perp ^{2}}\right) \frac{S_{\perp }}{\pi ^{2}\alpha _{s}}\frac{N_{c}^{2}-1}{N_{c}}%
\frac{1}{%
b_{\perp }^{2}\ln \frac{1}{b_{\perp }^{2}\Lambda ^{2}}}\left[ 1-\exp \left( -%
\frac{1}{4}b_{\perp }^{2}Q_{sg}^{2}\right) \right] \ .
\end{equation}
This can be compared to the normal Weizs\"{a}cker-Williams gluon distribution in $b_{\perp}$ space defined as $x\tilde{G}^{(1)}(x,b_\perp)=\int \text{d}^{2}q_\perp e^{-iq_\perp \cdot b_\perp} xG^{(1)}(x,q_\perp)$,
\begin{eqnarray}
x\tilde{G}^{(1)}(x,b_\perp)=\frac{S_{\perp }}{\pi ^{2}\alpha _{s}}\frac{N_{c}^{2}-1}{N_{c}}%
\frac{\ln \frac{1}{b_{\perp }^{2}\Lambda ^{2}}-2}{b_{\perp }^{2}\ln \frac{1}{b_{\perp }^{2}\Lambda ^{2}}}\left[ 1-\exp \left( -%
\frac{1}{4}b_{\perp }^{2}Q_{sg}^{2}\right) \right].
\end{eqnarray}
At small $b_\perp$, $x\tilde{h}_{\perp}^{(1)ij}(x,b_\perp)$ is proportional to
$\left(\delta^{ij}- {2b_\perp^i b_\perp^{j}}/{b_\perp ^{2}}\right)$ times a
constant, whereas $xG^{(1)}(x,b_\perp)$ behaves as $\ln \frac{1}{\Lambda^2b_\perp^2}$
due to the logarithmic term in $Q_{sg}^2$. These properties are
consistent with their perturbative behaviors at large transverse momentum~\cite{Sun:2011iw}.

Similarly for the dipole gluon counterparts, one gets
\begin{equation}
x\tilde{h}_{\perp}^{(2)ij}(x,b_\perp) = \frac{1}{2}\left(\delta_\perp^{ij}- \frac{2b_\perp^i b_\perp^{j}}{%
b_\perp ^{2}}\right)\frac{N_c S_\perp}{2\pi^2 \alpha_s} \exp[-\frac{1}{4}Q_{sq}^2b_\perp^2]Q_{sq}^2 \left[\frac{1}{\ln\frac{1}{\Lambda^2b_\perp^2}}+
\frac{b_\perp^2Q_{sq}^2}{4}\left(1-\frac{1}{\ln\frac{1}{\Lambda^2b_\perp^2}}\right)^2\right]\ ,
\end{equation}
and 
\begin{equation}
x\tilde{G}^{(2)}(x,b_\perp) =\frac{N_c S_\perp}{2\pi^2 \alpha_s} \exp[-\frac{1}{4}Q_{sq}^2b_\perp^2] Q_{sq}^2\left[1-\frac{2}{\ln\frac{1}{\Lambda^2b_\perp^2}}
-\frac{b_\perp^2Q_{sq}^2}{4}\left(1-\frac{1}{\ln\frac{1}{\Lambda^2b_\perp^2}}\right)^2\right]\ .
\end{equation}
Again, in the small $b_\perp$ limit, they behave the same as those Weizs\"{a}cker-Williams gluon
distributions, respectively. It is interesting to notice that their large $b_\perp$ behaviors
are different. For the dipole gluon distributions, they decrease exponentially whereas the
Weizs\"{a}cker-Williams ones have power behaviors.
These expressions can be viewed as the initial conditions of the
resummation discussed in Ref.~\cite{Sun:2011iw}.

\section{Small-$x$ evolution of the linearly polarized gluon distributions}

In this section, we discuss the small-$x$ evolution of the linearly
polarized gluon distributions. We separate the discussions into two parts:
the first part is on the evolution of the linearly polarized dipole gluon
distribution since it is trivial and it only involves the dipole amplitude;
then we derive the evolution equation for the linearly polarized Weizs\"{a}cker-Williams gluon distribution from the small-$x$ evolution equation of
quadrupoles.

\subsection{The evolution of the linearly polarized dipole gluon distribution}

According to the definition of the linearly polarized dipole gluon
distribution, and the above calculation of the cross section of dijet ($q\gamma^{\ast}$) production in Drell-Yan processes in pA collisions, we know
that the linearly polarized partner of the dipole gluon distribution is
identical to the normal dipole gluon distribution, i.e., $xh^{(2)}_{\perp}(x,q_\perp)=xG^{(2)}(x,q_\perp)$. In general, one can write
these distributions in terms of the dipole amplitude, namely, the two point function of
Wilson lines $\frac{1}{N_c}\left\langle\text{Tr}\left(U(x_{\perp})U^{\dagger}(y_{\perp})\right)\right\rangle$ as follows
\begin{equation}
xh^{(2)}_{\perp}(x,q_\perp)=xG^{(2)}(x,q_\perp)= \frac{q_{\perp }^2N_{c}}{
2\pi^2 \alpha_s}\int d^2 x_{\perp}\int \frac{d^2y_\perp}{(2\pi)^2}
e^{-iq_\perp\cdot (x_\perp-y_\perp)} \frac{1}{N_c}\left\langle\text{Tr}
U(x_{\perp})U^\dagger(y_\perp)\right\rangle_{Y} .
\end{equation}
The small-$x$ evolution of the dipole amplitude follows the well-known
Balitsky-Kovchegov equation \cite{Balitsky:1995ub+X,Kovchegov:1999yj} which reads
\begin{eqnarray}
\frac{\partial}{\partial Y}\left\langle\text{Tr}\left[U(x)U^{\dagger}(y)
\right]\right\rangle_{Y}&=&-\frac{\alpha_s N_c}{2\pi^2}\int d^2 z_{\perp}
\frac{(x_{\perp}-y_{\perp})^2}{(x_{\perp}-z_{\perp})^2(z_{\perp}-y_{\perp})^2
}  \notag \\
&&\times \left\{ \left\langle\text{Tr}\left[U(x)U^{\dagger}(y)\right]
\right\rangle_{Y}-\frac{1}{N_c}\left\langle\text{Tr}\left[U(x)U^{\dagger}(z)
\right]\text{Tr}\left[U(z)U^{\dagger}(y)\right]\right\rangle_{Y}\right\}.
\label{bk}
\end{eqnarray}
In the dilute regime, the Balitsky-Kovchegov equation reduces to the famous
BFKL equation which leads to the exponential growth in terms of the rapidity
$Y\simeq \ln \frac{1}{x}$.

\subsection{The evolution of the linearly polarized Weizs\"{a}cker-Williams gluon distribution}

The operator definition of the Weizs\"{a}cker-Williams gluon distribution can be obtained from the
quadrupole correlator whose initial condition can be throughly calculated in the MV model. In Refs.~\cite{JalilianMarian:2004da,Dominguez:2011gc, Dumitru:2010ak,Iancu:2011ns}, the small-$x$ evolution equation of the quadrupole has been derived and studied analytically. Similarly to the Balitsky-Kochegov equation for dipoles, quadrupoles follow BFKL evolution in the dilute regime and reach the saturation regime as a stable fixed point. In addition, one expects that quadrupoles should also exhibit the same geometrical scaling behavior as dipoles. Recently, using the JIMWLK renormalization equation \cite{Jalilian-Marian:1997jx+X, Ferreiro:2001qy}, the first numerical studies \cite{Dumitru:2011vk} of the small-$x$ evolution of quadrupoles indeed observe evidence of traveling wave solutions and geometric scaling for the quadrupole. According to \cite{Metz:2011wb} and Ref.~\cite{Dominguez:2010xd, Dominguez:2011wm, Bomhof:2006dp}, the Weizs\"{a}cker-Williams gluon distribution can be written as
\begin{eqnarray}
xG^{ij}_{\text{WW}}(x,k_\perp)&=&-\frac{2}{\alpha_S}\int\frac{d^2v}{(2\pi)^2}
\frac{d^2v^{\prime}}{(2\pi)^2}\;e^{-ik_\perp\cdot(v-v^{\prime})}\left\langle
\text{Tr}\left[\partial^iU(v)\right]U^\dagger(v^{\prime})\left[
\partial^jU(v^{\prime})\right]U^\dagger(v)\right\rangle_{Y}. \\
&=&\frac{\delta^{ij}}{2}xG^{(1)}(x,q_{\perp })+\frac{1}{2}\left(2
\frac{q_{\perp }^iq_{\perp }^j}{q_{\perp}^2}-\delta^{ij}\right)xh^{(1)}_{
\perp}(x,q_{\perp }).
\end{eqnarray}
The evolution equation for the correlator $\left\langle\text{Tr}\left[
\partial^iU(v)\right]U^\dagger(v^{\prime})\left[\partial^j U(v^{\prime})
\right]U^\dagger(v)\right\rangle_{Y}$ can be obtained from the evolution equation of the quadrupole $\frac{1}{N_c}\left\langle\text{Tr}\left(U(x_{1})U^{
\dagger}(x^{\prime}_{1})U(x_{2})U^{\dagger}(x^{\prime}_{2})\right)\right\rangle_{Y}$
by differentiating with respect to $x_1^i$ and $x_2^j$, and then setting $x^i_1=x_2^{\prime i}=v^i$ and $x^j_2=x_1^{\prime j}=v^{\prime j}$. Then the
resulting evolution equation becomes\footnote{This evolution equation involves derivatives of the Wilson lines and complicated kernels which make it very hard to solve directly. However, one can extract the evolution information by numerically solving the evolution equation for quadrupoles first and then making numerical differentiation and identification of coordinates.}
\begin{eqnarray}
&&\frac{\partial}{\partial Y}\left\langle\text{Tr}\left[\partial^iU(v)\right]
U^\dagger(v^{\prime})\left[\partial^j U(v^{\prime})\right]
U^\dagger(v)\right\rangle_{Y}  \notag \\
&=&-\frac{\alpha_s N_c}{2\pi^2}\int d^2 z_{\perp} \frac{(v-v^\prime)^2}{
(v-z)^2(z-v^\prime)^2}\left\langle\text{Tr}\left[\partial^iU(v)\right]
U^\dagger(v^{\prime})\left[\partial^jU(v^{\prime})\right]U^\dagger(v)\right
\rangle_{Y}  \notag \\
&&-\frac{\alpha_s N_c}{2\pi^2}\int d^2 z_{\perp} \frac{1}{N_c}\frac{
(v-v^\prime)^2}{(v-z)^2(z-v^\prime)^2}\left[\frac{(v-v^\prime)^i}{
(v-v^\prime)^2}-\frac{(v-z)^i}{(v-z)^2}\right]  \notag \\
&&\quad \times \left\{\left\langle\text{Tr}\left[U(v)U^\dagger(v^{\prime})
\left[\partial^jU(v^{\prime})\right]U^\dagger(z)\right]\text{Tr}\left[
U(z)U^\dagger(v)\right]\right\rangle_{Y}-\left\langle\text{Tr}\left[
U(z)U^\dagger(v^{\prime})\left[\partial^jU(v^{\prime})\right]U^\dagger(v)
\right]\text{Tr}\left[U(v)U^\dagger(z)\right]\right\rangle_{Y} \right\}
\notag \\
&&-\frac{\alpha_s N_c}{2\pi^2}\int d^2 z_{\perp} \frac{1}{N_c}\frac{
(v-v^\prime)^2}{(v-z)^2(z-v^\prime)^2}\left[\frac{(v^\prime-v)^j}{
(v^\prime-v)^2}-\frac{(v^\prime-z)^j}{(v^\prime-z)^2}\right]  \notag \\
&&\quad \times \left\{\left\langle\text{Tr}\left[\left[\partial^iU(v)\right]
U^\dagger(z)U(v^\prime)U^\dagger(v)\right]\text{Tr}\left[U(z)U^\dagger(v^
\prime)\right]\right\rangle_{Y}-\left\langle\text{Tr}\left[\left[
\partial^iU(v)\right]U^\dagger(v^\prime)U(z)U^\dagger(v)\right]\text{Tr}
\left[U(v^\prime)U^\dagger(z)\right]\right\rangle_{Y} \right\}  \notag \\
&&-\frac{\alpha_s N_c}{4\pi^2}\int d^2 z_{\perp} \frac{1}{N_c}\left[
\partial^i_v \partial^j_{v^{\prime}}\frac{(v-v^\prime)^2}{(v-z)^2(z-v^\prime)^2
}\right]  \notag \\
&&\quad \times \left\{\left\langle\text{Tr}\left[U(v^\prime)U^{\dagger}(z)
\right]\text{Tr}\left[U(z)U^{\dagger}(v^\prime)\right]\right\rangle_{Y}+
\left\langle\text{Tr}\left[U(v)U^{\dagger}(z)\right]\text{Tr}\left[
U(z)U^{\dagger}(v)\right]\right\rangle_{Y}\right .  \notag \\
&&\quad \quad \left .-\left\langle\text{Tr}\left[U(v^\prime)U^{\dagger}(v)
\right]\text{Tr}\left[U(v)U^{\dagger}(v^\prime)\right]\right
\rangle_{Y}-N_c^2\right\}.  \label{wwe}
\end{eqnarray}

The evolution equation of the Weizs\"{a}cker-Williams
gluon distributions can be obtained by contracting the above correlator with $\delta_{ij}$ and the one for the linearly polarized partner by contracting with $\left(2%
\frac{q_{\perp }^iq_{\perp }^j}{q_{\perp}^2}-\delta^{ij}\right)$. Although the expression is quite complicated in general, the result gets simplified in the dilute regime as in Ref.~\cite{Dominguez:2011gc}. In the dilute regime, the correlator which yields the Weizs\"{a}cker-Williams
gluon distribution can be reduced to a simple form in terms of $\Gamma(v,v^\prime)$
\begin{equation}
\left\langle\text{Tr}\left[\partial^iU(v)\right]%
U^\dagger(v^{\prime})\left[\partial^jU(v^{\prime})\right]U^\dagger(v)\right%
\rangle_{Y}
=\frac{C_F}{2}\partial_v^i\partial_{v^\prime}^j \Gamma(v,v^\prime)_Y, \label{w2}
\end{equation}
where $\frac{C_F}{2} \Gamma(v,v^\prime)$ is the leading order dipole amplitude which satisfies the BFKL equation
\begin{equation}
\frac{\partial}{\partial Y}\Gamma(x_1,x_2)_Y=\frac{N_c\alpha_s}{2\pi^2}\int d^2z\,\frac{(x_1-x_2)^2}{(x_1-z)^2(x_2-z)^2}\left[\Gamma(x_1,z)_Y+\Gamma(z,x_2)_Y-\Gamma(x_1,x_2)_Y\right].\label{bfkl}
\end{equation}

In the dilute regime where the gluon density is low, we know that the Weizs\"{a}cker-Williams gluon distributions, $xG^{(1)}(x,q_{\perp })$ and $xh^{(1)}_{\perp}(x,q_{\perp })$, as well as the dipole gluon distributions all reduce to the same leading twist result. Therefore, despite of the distinct behavior in the saturation regime, we find that all these four types of gluon distributions follow the BFKL equation in the dilute regime where the gluon density is low. The physical consequence of this results is that the linearly polarized gluon distributions should be as important as the normal gluon distributions in the low-$x$ region since they also receive the exponential rise in rapidity $Y$ due to the BFKL evolution. Furthermore, according to the discussion in Ref.~\cite{Munier:2003vc,Mueller:2002zm}, the BFKL evolution together with a saturation boundary can give rise to the geometrical scaling behavior \cite{Stasto:2000er,GolecBiernat:2001if,Iancu:2002tr} of the dipole gluon distribution. Since the quadrupole evolution equation also contains the same property as discussed in Ref.~\cite{Dominguez:2011gc, Dumitru:2011vk}, the Weizs\"{a}cker-Williams gluon distribution and its linearly polarized partner should exhibit geometrical scaling behavior as well, although their evolution equations are much more complicated in the saturation regime. In terms of the traveling wave picture \cite{Braun:2000wr, Munier:2003vc} for the evolution of dipoles and quadrupoles, the velocities of the traveling waves for dipoles and quadruples are identical, since the velocity is determined by BFKL evolution. This implies that the energy dependence of the saturation momentum $Q_s^2\simeq Q_0^2 (x_0/x)^{\lambda}$ with $Q_0=1 GeV$, $x_0=3\times 10^{-3}$ and $\lambda=0.29$,  should be universal for all these four different gluon distributions.

\section{Conclusion}
 We perform the color dipole model calculation of the cross section of DIS dijet and Drell-Yan dijet processes, and demonstrate that the linearly polarized partners of the Weizs\"{a}cker-Williams and dipole gluon distributions naturally arise in these processes. This result is in complete agreement with Ref.~\cite{Metz:2011wb} and implies that the measurement of the $\cos \left(2\Delta \phi\right)$ asymmetries in these dijet processes can be a direct probe of these two different linearly polarized gluon distributions. In addition, the small-$x$ evolution studies of the linearly polarized gluon distributions reveals that they also rise exponentially as function of the rapidity at high energy and they should also exhibit the geometrical scaling behavior as the normal unpolarized gluon distributions do.

\begin{acknowledgements}
We thank Prof. A. H. Mueller for helpful discussions. This work was supported in part by the U.S. Department of Energy under the DOE OJI grant No. DE - SC0002145 and contract number DE-AC02-98CH10886.
\end{acknowledgements}

\end{document}